\documentclass[aps,twocolumn,showpacs]{revtex4}
\usepackage{graphicx}
\usepackage{bm}
\usepackage{amsmath}

\begin{document}

\title{Resonant intrinsic spin Hall effect in p-type GaAs quantum well
structure}
\author{Xi Dai$^{1,2,3}$, Zhong Fang$^3$, Yu-Gui Yao$^3$ and Fu-Chun Zhang$^{1,2}$}
\affiliation{$^1$ Department of Physics, the University of Hong Kong, Hong Kong\\
$^2$ Center of Theoretical and Computational Physics, the University of Hong
Kong, Hong Kong\\
$^3$ Institute of Physics, Chinese Academy of Sciences, Beijing, China }
\date{\today}

\begin{abstract}
We study intrinsic spin Hall effect in p-type GaAs quantum well structure
described by Luttinger Hamiltonian and a Rashba spin-orbit coupling arising
from the structural inversion symmetry breaking. The Rashba term induces an
energy level crossing in the lowest heavy hole subband, which gives rise to
a resonant spin Hall conductance. The resonance may be used to identify the
intrinsic spin Hall effect in experiments.
\end{abstract}


\maketitle

\narrowtext

The study of spin Hall effect (SHE), in which an electric field
induces a transverse spin current, has recently evolved into a
subject of intense research for its potential application to the
information processing. The intrinsic SHE was proposed by Murakami
et al.\cite{murakami} in
p-type semiconductor of a Luttinger Hamiltonian and by Sinova et al.\cite%
{sinova} in 2-dimensional (2D) electron systems with Rashba spin-orbit
coupling. Their works have generated a lot of theoretical activities.~\cite%
{culcer,schliemann,hu,shen,rashba,
inoue,shenprl,burkov,mishchenko,khaetskii,raimondi,
chalaev,loss,bernevig2,murakami2,sheng,macdonald} Current
theoretical understanding is that the intrinsic SHE does not
survive in the diffusive transport in the thermodynamic limit for
the 2D Rashba electron
system~\cite{rashba60} in the absence of strong magnetic fields ~\cite%
{inoue,mishchenko,khaetskii,raimondi,chalaev}, but the effect
appears to be robust in the 2D hole gases~\cite{loss,bernevig2},
p-doped bulk semiconductors and the modified Rashba coupling
case~\cite{murakami2}. The earlier theoretical work on the
extrinsic SHE is associated with the impurity scattering, such as
the skew scattering and the side jump processes
~\cite{dyakonov,hirsch,szhang}. On the experimental side, there
have been two groups reporting the observation of spin Hall
effect. Kato et al.~\cite{kato} used Kerr rotation microscopy to
detect and image electrically induced electron-spin polarization
near the edge of a n-type semiconductor channel. The effect was
suggested to be extrinsic based on the weak dependence on crystal
orientation for the strained samples. Wunderlich et al.
~\cite{wunderlich} observed the SHE in 2D hole system with
spin-orbit coupling, and interpreted the effect to be intrinsic.
In view of the unfamilarity of the spin Hall transport, it will be
desirable and important to experimentally identify if the observed
SHE is intrinsic. Such an identification requires careful study of
properties of the intrinsic SHE.

In this Letter we study the intrinsic SHE in p-type GaAs quantum
well structure described by a Luttinger Hamiltonian with a Rashba
spin-orbit coupling arising from the structural inversion symmetry
breaking. The Rashba term hybridizes the electronic sub-bands
of the Luttinger Hamiltonian in a quantum well and induces energy
level crossings in the both heavy and light hole sub-bands. The level
crossing, if it occurs at the Fermi level, gives rise to a
resonant intrinsic SHE characterized by a sharp peak and a sign
change in the spin Hall conductance. By tuning the Rashba coupling
strength and/or carrier density, this type of resonance may be
observed in experiment to distinguish the intrinsic from the
extrinsic SHE. In the latter case, one does not expect such
drastic changes.  The sign of the extrinsic SHE induced in the skew scattering,
which dominates over the side jump process in the weak disorder
limit of our interest here, depends on the sign of the impurity
potential, and does not change with changing carrier density or the
Rashba coupling strength.

We consider an effective Hamiltonian for the hole doped quantum well with
the structural inversion symmetry breaking, described by the Luttinger
Hamiltonian with a confinement potential along the z direction and an
additional Rashba coupling term,
\begin{equation}
H=H_{L}-\lambda (\hat{z}\times \vec{p} ) \cdot \vec{S}+ V(z)  \label{H}
\end{equation}
where $\lambda$ is the Rashba spin-orbit coupling, $\vec p$ is the momentum,
and $\vec S = (S_x, S_y, S_z)$ are the spin-3/2 operators. $V(z)$ is a
confinement potential along the z-direction. For simplicity, we choose $%
V(z)=+\infty$ for $|z|>L$ and $V(z)=0$ otherwise. Note that we
have assumed that the only effect of the structural inversion
symmetry breaking is to induce a Rashba coupling term in Eq. (1),
and any asymmetry in $V(z)$ has been neglected. We expect that the
simplification of $V(z)$ will not alter the qualitative physics we
discuss below. $H_{L}$ is the Luttinger effective Hamiltonian
describing the hole motion in the valence band,
\begin{equation}
H_{L}=- \frac{\hbar ^{2}}{2m} \left( (\gamma _{1}+\frac{5\gamma _{2}}{2})
\bigtriangledown ^{2} + 2\gamma _{2}(\vec S \cdot \vec{\bigtriangledown}%
)^{2} \right)  \label{H_L}
\end{equation}
Since the translational symmetry is broken only along the z direction, the
momentum $\hbar \vec k$ in the $x-y$ plane remains to be a good quantum
number. For a given $\vec k$, $H$ can be reduced to a 1D effective
Hamiltonian
\begin{eqnarray}
H_{\vec k} = \frac{\hbar^2}{2m}(k^2 -\partial_{z}^2) (\gamma _{1}+ \frac{%
5\gamma _2}{2}) + V(z)  \notag \\
-\frac{\hbar ^{2}\gamma _{2}}{m} (S_x k_x +S_y k_y - S_{z}i\partial
_{z})^{2} -\lambda \hbar (k_y S_x -k_x S_y) ,  \label{Hk}
\end{eqnarray}

In the special case $k=0$, $S_{z}$ is a good quantum number and the eigen
wave-functions of $H_{\vec k}$ are found to be,
\begin{eqnarray}
\Psi _{\alpha n}(z)&=&\cos{(q_{n}z)}\chi _{a}, \, n=odd  \notag \\
\Psi _{\alpha n}(z)&=&\sin{(q_{n}z)}\chi _{a}, \, n=even.  \label{basis1}
\end{eqnarray}
with $q_{n}=n\pi/2L$, and $n$ being positive integers. $\chi _{a}$
is the eigenstate of $S_{z}$ corresponding to the eigenvalue of
$S_z =a=3/2,-3/2,1/2,-1/2$. The eigenstates are two-fold
degenerate corresponding to the eigenvalues
\begin{eqnarray}
E_{\pm 3/2,n}=\hbar ^{2}q_{n}^{2}/2m_{hh}  \notag \\
E_{\pm 1/2,n}=\hbar ^{2}q_{n}^2/2m_{lh},
\end{eqnarray}
with $m_{lh}=m/(\gamma _{1}+2\gamma _{2})$ and $m_{hh}=m/(\gamma
_{1}-2\gamma _{2})$ the effective masses for light and heavy hole subbands,
respectively. The splitting of the heavy and light hole subbands at $k=0$ is
due to the $\gamma_2$ term in $H_L$. Note that the Rashba coupling term
vanishes at $k=0$.

For $\vec k \neq 0$, $S_z$ is no longer a good quantum number, and the
two-fold degeneracy splits and all the heavy and light hole subbands mix to
each other in general. Two limiting cases were considered previously. One is
the limit $2k L/\pi \ll 1$, while the Rashba coupling $\lambda$ is finite.
This case was studied by Schliemann and Loss\cite{schliemann}, who used a
lowest order perturbation theory to derive a simplified effective theory by
approximately projecting the full Hamitionian to the lowest heavy hole
subband. The splitting of the lowest heavy hole subband due to the Rashba
coupling was found to be proportional to $k^{3}$, which gives the value of
the spin Hall conductance of the order of $9e/8\pi$. In this limit, the spin
Hall effect is purely contributed from the Rashba term. The other limiting
case is $\lambda=0$, which was considered by Bernevig and Zhang~\cite%
{bernevig2}, who calculated the spin Hall conductance by including both the
lowest heavy hole and light hole subbands. The spin Hall effect in this case
is purely caused by the Luttinger type spin-orbit coupling.

Below we shall study the electronic structure of Eq. (1) at a
finite $\vec{k} $ and $\lambda $. We use basis wavefunctions of
the $k=0$ eigenstates, given by Eqs. (4), and apply a truncated
method, in which only $N$ basis states of the lowest energies of
Eqs. (5) are kept. We then diagonize $H_{\vec{k}}$ within this
truncated Hilbert space by numerical means. As $N$ increases, the
eigen energies of the lowest subbands converge quickly. In Fig.
1(a) we plot the lowest four subbands in the Rashba free case,
namely HH1,LH1,HH2 and HH3 from the bottom to top, with a double
degeneracy for each subband. Here HH and LH denote heavy hole and
light hole respectively. In our calculations, we use $m=\,\gamma
_{1}=7.0,\,\gamma _{2}=1.9$. With this choice of the parameters,
the correct band structure of the sub-bands are reproduced~\cite{footnote},
and the results
are in good agreement with the previous calculations using
the evolope function method~\cite{huang,russian}. The Rashba term
lifts the double degeneracy of each su-bband at finite $k$, as
shown in Fig. 1(b) and (c). For $k\ll \pi /2L$, the energy
splitting is found to be proportional to $k^{3}$ for the HH1 band
and to $k$ for the LH1 subband, consistent with the previous
study based on the leading order purturbation around the $\Gamma $ point~%
\cite{schliemann}. With the increment of $k$, the interplay between the
Rashba and Luttinger type spin-orbit couplings leads to a level crossing
within the sub-bands. For relatively small Rashba coupling ($\lambda =\hbar
^{2}/mL)$, the level crossing occurs only in the LH1 subband. While for
large Rashba coupling ($\lambda =3\hbar ^{2}/mL)$, level crossings are found
in both LH1 and HH1 subbands. A careful analysis reveals that
HH2 subband is important to the level crossings in both HH1 and LH1 subbands.

\begin{figure}[tbp]
\begin{center}
\includegraphics[width=8cm,angle=0,clip=]{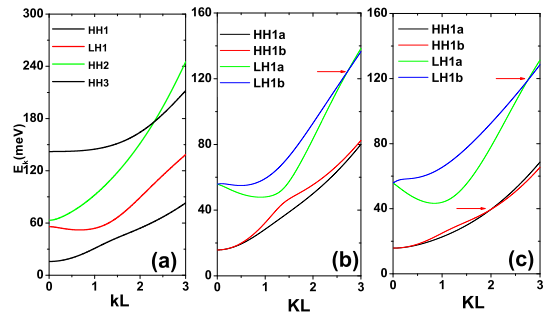}
\end{center}
\caption{Dispertion of low lying
subbands for p-type GaAs quantum well struction with the width $2L=83%
\mathring{A}$. (a): for Rashba coupling $\protect\lambda =0$, (b): $\protect%
\lambda =\hbar ^{2}/mL$, and (c): $\protect\lambda =3.0\times
\hbar ^{2}/mL$. } \label{fig1}
\end{figure}

We now discuss the spin Hall conductance. We consider a linear response of
the spin current tensor component $j_{s, x}^z$ to a transverse electric
field along the $y$-direction, where we define $\ j_{s,x}^{z}= (v_{x}S_{z}+
S_z v_x)/2$, and $v_{\alpha}= \partial H/\partial p_{\alpha}$ is the $\alpha
=x-, \, y-$ component of the velocity operator. The spin Hall conductance
can be calculated by using Kubo formula,

\begin{eqnarray}
\sigma _{x}^{z} &=&-\frac{2e}{\hbar }\int \frac{d^{2}\vec{k}}{(2\pi )^{2}}%
\sum_{i^{\prime }>n}\frac{f(E_{i^{\prime }\vec{k}})-f(E_{i\vec{k}})}{%
(E_{i^{\prime }\vec{k}}-E_{i\vec{k}})^{2}}  \notag \\
&&\times Im\left\langle i\left\vert j_{s,x}^{z}\right\vert i^{\prime
}\right\rangle \left\langle i^{\prime }\left\vert v_{y}\right\vert
i\right\rangle   \label{kubo}
\end{eqnarray}%
where $f$ is the Fermi distribution function, and $E_{i\vec{k}}$ is the
energy of the $i^{th}$ subband with the in-plane momentum $\vec{k}$. The
calculated spin Hall conductance at zero temperature as a function of $%
\lambda $ and hole density for GaAs quantum well is shown in Fig. 2 and Fig. 4,
where we have assumed a carrier lifetime $\tau =2.0\times 10^{-11}s$.
At a lower hole density and a large $\lambda$, there is a resonance
associated with the level crossing of the HH1 subbands.
At a higher hole density, the resonance is associated
with the level crossing of the LH1 subbands, insensitive to the
value of $\lambda $. In Fig. 2 we show $\sigma _{x}^{z}$ as a function of $%
\lambda $ for a lower hole density case. A resonance is clearly seen at $%
\lambda \approx 3.15 \hbar ^{2}/mL$, associated with the level crossing of the
HH1 subbands at the Fermi energy. The resonance becomes a singularity in $%
\sigma _{x}^{z}$ if we use $\tau \rightarrow \infty $ and is smoothen out if
$\tau $ is 10 times smaller. The resonance may be used to identify the
intrinsic SHE by tuning the Rashba coupling in experiments.

\begin{figure}[tbp]
\begin{center}
\includegraphics[width=8cm,angle=0,clip=]{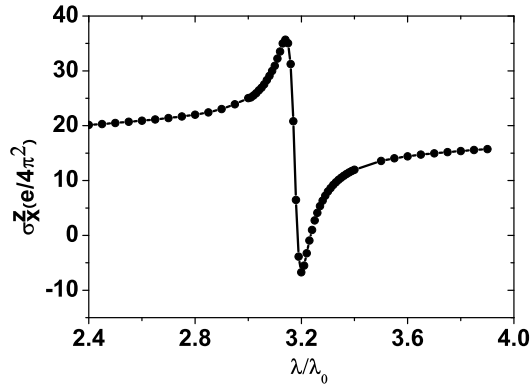}
\end{center}
 \caption{Spin Hall conductance of GaAs quantum well
as a function of dimensionless Rashba coupling $\lambda/\lambda_0$, with
$\lambda_0=\hbar ^{2}/mL$ . The hole density $n_{p}=5.0\times
10^{11}/cm^{2}$ and half-thickness $2L=83\mathring{A}$. A finite life time $\protect\tau %
=2.0\times 10^{-11}s$, equivalent to a mobility of $10^{4}cm^2/sV$, is
assumed. }
 \label{fig2}
\end{figure}

In Fig. 3 we plot the resonant Rashba coupling associated with the level
crossing of the HH1 subbands as functions of hole density in GaAs quantum
well for various well thickness. The parameters corresponding to the resonance
appears to be within the experimental region.

\begin{figure}[tbp]
\begin{center}
\includegraphics[width=8cm,angle=0,clip=]{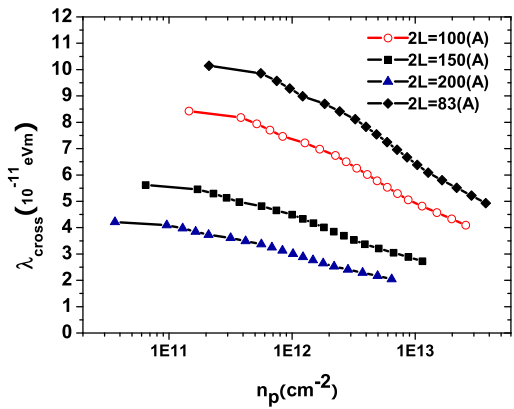}
\end{center}
\caption{ The resonant Rashba coupling for the HH1 subbands as
functions of hole density for various thickness $2L$ in GaAs
quantum well.}
 \label{fig3}
\end{figure}

A few remarks are in order. The first is on the possible cancellation of the
SHE due to the vortex correction. Recent theoretical
works show that the vertex correction is severe in the 2D Rashba
electron system, but is discounted for the 2D hole gases and the p-type bulk
semiconductor ~\cite{loss,bernevig2,murakami2}. The resonance we
discuss here is related to the level crossing in the
heavy hole sub-bands in the Luttinger Hamiltonian with a Rashba coupling,
and the leading order in the energy splitting $\propto k^3$, very different
from the electron Rashba system where the energy splitting is $\propto k$.
We expect the resonance be robust against weak spin diffusion. Secondly,
the resonance predicted in this system resembles the level crossings in the
ferromagnetic metal with a magnetic monopole-like structure in the momentum space
~\cite{fang}, and in the 2D quantum Hall system with a Rashba coupling
~\cite{shenprl}.

In Fig. 4, we plot the spin Hall conductance as a function of hole
density for a weak Rashba coupling $\lambda =\hbar ^{2}/mL$. In
this case, the level crossing occurs only at the LH1 subbands, and
the resonance SHE appears at $n_{p}=1.2\times
10^{13}/cm^{2}$. As the carrier density or chemical potential
increases from zero, there are two step-like features in
$\sigma_x^z$ at $n_{p}=0$ and $n_{p}=4.0\times 10^{12}/cm^{2}$
corresponding to the bottoms of the HH1 and LH1 subbands
respectively, caused by the Rashba term. The spin Hall conductance
at $n_{p}\rightarrow 0$ is found to be around
$1.2e/\pi $, which agrees well with the results obtained by Schliemann
and Loss~\cite{schliemann}. The dip feature around
$n_{p}=2.5\times 10^{12}/cm^{2}$ is due to the negative effective
mass of the LH1 subband near the $\Gamma $ point. The dashed line
plotted in Fig. 4 is the spin Hall conductance at $\lambda=0$.
This shows that the resonance is contributed from the
interplay between the Rashba coupling and the spin-orbit coupling
in the Luttinger Hamiltonian.  In terms of the Kubo formula Eq. (6),
the resonance is contributed from the intra-subband
transition. The robustness of the resonance associated with the
LH1 subband level crossing needs more causious.
The energy splitting of the LH1 subbands is $\propto \lambda k$ at
small $k$, similar to that in the Rashba electron system. Near the
level crossing point, the term $\lambda k^3$ also becomes
important. Without the $k^3$ term, the intronsic SHE does not survive.
It might be possible that the $k^3$ term discounts the
vortex effect so that the SHE survives. However, this will need
further study.

In the experiment of Wunderlich et al.~\cite{wunderlich}, the hole
density is $2.0\times 10^{12}cm^{-2}$ and the effective width of
the quantum well can be estimated to be $2L=83\mathring{A}$ by
fitting the $\Gamma$ point subband splitting of the LH1 and HH1
subbands. The Rashba coupling constant can also be extracted by
fitting the splitting of the HH1 subband at the Fermi level, which
is approximately $\lambda=1.5\times 10^{-11}eVm$. From
figure\ref{fig3}, the required Rashba coupling for the resonance
is around $8.5\times 10^{-11}eVm$, which is several times larger
than the parameter in Wunderlich's experiment. In order to observe
the resonance in the HH1 subband, one will need to either increase
the Rashba coupling by about 6 times or to increase the thickness
of the quantum well to around $200\mathring{A}$ while keeping the
2D carrier density unchanged. Note that as shown in
figure\ref{fig2}, the resonance requires quite high mobility
(around $10^4cm^2/sV$).  To observe the possible resonance
asscoiated with the level crossing in the LH1 subband, one would
need to tune the carrier density around $1.2\times 10^{13}cm^{-2}$
or the well thickness. The value of the Rashba coupling does not
play much role in this case.

\begin{figure}[htb]
\begin{center}
\includegraphics[width=8cm,angle=0,clip=]{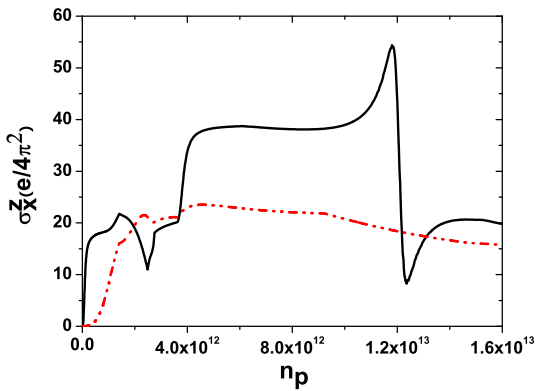}
\end{center}
\caption{Spin Hall conductance as a function of hole density for
GaAs quantum well of half-thickness $2L=83 \AA$ and Rashba
coupling $\lambda =\hbar ^2/mL$. Dashed line is
the spin Hall conductance at $\lambda=0$.} \label{fig4}
\end{figure}

In summary, we have studied the electronic structure and the
intrinsic transverse spin transport properties of the p-type GaAs
quantum well. The Rashba spin-orbit coupling arising from the
structure inversion symmetry breaking splits the subbands of the
Luttinger Hamiltonian, and induces level
crossings within the lowest heavy hole subbands and the lowest
light hole subbands. These level crossings, if occuring at the
Fermi level, give rise to resonant spin Hall conductance. Our
calculations show that the parameters (the hole density, the well
thinkness, and the Rashba coupling strength) for the resonance are
likely reachable in experiments. This phenomenon can be used to
distinguish the intrinsic spin Hall effect from the extrinsic one.
We expect the resonant spin Hall effect associated with the heavy
hole subbands be robust since the vertex correction to the heavy
hole subband at small momentum has been shown to be non-severe.
The robustness of the resonant effect associated with the light hole sub-bands
requires careful examination of the vertex corrections. We
have been benefited from many useful and stimulating discussions
with Bradley Foreman, Shun-Qing Shen, Zidan Wang, Shou-Cheng Zhang, to
whom we would like to thank. This work was supported in part by
Hong Kong's RGC grant and NSFC in China.

\end{document}